\documentclass[12pt]{article}
\usepackage[round,comma,numbers,sort&compress]{natbib}
\usepackage{graphicx}
\usepackage{float}
\usepackage[caption = false]{subfig}
\usepackage[a4paper,margin=1in]{geometry}
\usepackage{times}
\usepackage{amsmath}
\usepackage{xspace}
\usepackage{amssymb}
\usepackage{xr}
\usepackage[]{changes}
\definecolor{DarkGreen}{rgb}{0.0,0.4,0.0}  

\newcommand{\spacing}[1]{\renewcommand{\baselinestretch}{#1}\large\normalsize}
\spacing{1.25}

\newenvironment{affiliations}{%
    \setcounter{enumi}{1}%
    \setlength{\parindent}{0in}%
    \slshape\sloppy%
    \begin{list}{\upshape$^{\arabic{enumi}}$}{%
        \usecounter{enumi}%
        \setlength{\leftmargin}{0in}%
        \setlength{\topsep}{0in}%
        \setlength{\labelsep}{0in}%
        \setlength{\labelwidth}{0in}%
        \setlength{\listparindent}{0in}%
        \setlength{\itemsep}{0ex}%
        \setlength{\parsep}{0in}%
        }
    }{\end{list}\par\vspace{12pt}}

\renewenvironment{abstract}{%
    \setlength{\parindent}{0in}%
    \setlength{\parskip}{0in}%
    \bfseries%
    }{}

\newenvironment{addendum}{%
    \setlength{\parindent}{0in}%
    \small%
    \begin{list}{Acknowledgements}{%
        \setlength{\leftmargin}{0in}%
        \setlength{\listparindent}{0in}%
        \setlength{\labelsep}{0em}%
        \setlength{\labelwidth}{0in}%
        \setlength{\itemsep}{12pt}%
        }
    }
    {\end{list}\normalsize}

\newcommand{\suppmv}{Supplementary Movie}

\newcommand{\suppnote}{Supplementary Notes\xspace}
\newcommand{\ion}[2]{#1$\;${\small\rmfamily\rom{#2}}\relax}%
\newcommand*{\rom}[1]{\uppercase\expandafter{\romannumeral #1\relax}}
\newcommand{\kms}{~km~s$^{-1}$\xspace} 

\newcommand{\aap}{    {\it Astron. Astrophys.}}

\newcommand{\apj}{    {\it Astrophys. J.}}
\newcommand{\apjl}{   {\it Astrophys. J. Lett.}}

\newcommand{\apss}{   {\it Astrophys. Space Sci.}}

\newcommand{\jgr}{    {\it J. Geophys. Res.}}
\newcommand{\mnras}{  {\it Mon. Not. Roy. Astron. Soc.}}
\newcommand{\nat}{    {\it Nature}}

\newcommand{\solphys}{{\it Solar Phys.}}
\newcommand{\ssr}{    {\it Space Sci. Rev.}}

\title{The Birth of A Coronal Mass Ejection} 
\date{\vspace{-5ex}}
\author{Tingyu Gou$^{1,2}$, Rui Liu$^{1}$, Bernhard Kliem$^{3,4}$, Yuming Wang$^{1,5}$, Astrid M. Veronig$^{2}$}  

\begin{document}

\maketitle
\begin{affiliations}
	\item CAS Key Laboratory of Geospace Environment, Department of Geophysics and Planetary Sciences, University of Science and Technology of China, Hefei 230026, China; rliu@ustc.edu.cn
	\item Institute of Physics/IGAM \& Kanzelh\"{o}he Observatory, University of Graz, Universit\"{a}tsplatz 5, 8010 Graz, Austria
	\item Institute of Physics and Astronomy, University of Potsdam, 14476 Potsdam,
		Germany; bkliem@uni-potsdam.de
    \item Mullard Space Science Laboratory, University College London,
		Holmbury St.~Mary, Dorking, Surrey RH5 6NT, UK
	\item Synergetic Innovation Center of Quantum Information \& Quantum Physics, University of Science and Technology of China, Hefei 230026, China	
	
\end{affiliations}
\begin{abstract}
The Sun's atmosphere is frequently disrupted by coronal mass ejections (CMEs), coupled with flares and energetic particles. In the standard picture, the coupling is explained by magnetic reconnection at a vertical current sheet connecting the flare loops and the CME, with the latter embedding a helical magnetic structure known as flux rope. As it jumps upward due to instabilities or loss of equilibrium, the flux rope stretches the overlying coronal loops so that oppositely directed field is brought together underneath, creating the current sheet. However, both the origin of flux ropes and their nascent paths toward eruption remain elusive. Here we present an observation of how a stellar-sized CME bubble evolves continuously from plasmoids, mini flux ropes that are barely resolved, within half an hour. The eruption initiates when plasmoids springing from a vertical current sheet merge into a leading plasmoid occupying the upper tip of the current sheet. Rising at increasing speed to stretch the overlying loops, this leading plasmoid then expands impulsively into the CME bubble, in tandem with hard X-ray bursts. This observation illuminates for the first time a complete CME evolutionary path that has the capacity to accommodate a wide variety of plasma phenomena by bridging the gap between micro-scale dynamics and macro-scale activities.
\end{abstract}

\section*{Introduction}
The eruptions in the solar atmosphere exhibit distinctly diverse patterns across a vast range of spatio-temporal scales, from coronal mass ejections (CMEs) in the shape of stellar-sized bubbles, to localized flares within active regions harboring sunspots, to collimated jets down to the resolution limit of modern telescopes. Fifty years of studies on flares and CMEs have converged to a standard picture: a flux rope is destabilized and a current sheet develops underneath \cite{Svestka&Cliver1992}, where successive magnetic reconnections add layers of plasma and magnetic flux to the snowballing CME bubble and simultaneously produce flare loops beneath the current sheet \cite{Lin&Forbes2000,Lin2004}. However, the flux rope's origin remains elusive. It has long been debated whether the flux rope forms in a magnetically sheared arcade during the course of the eruption \citep{Antiochos1999,Karpen2012}, or, exists before eruption resulting either from a sub-photospheric rope emerging into the corona \cite{Fan2009,Archontis&Hood2012} or from reconnection between coronal field lines \cite{vanBallegooijen&Martens1989,Moore2001} driven by the gradually evolving photosphere \cite{Amari2014}. It is also debated how the eruption initiates \cite{Forbes2006}, whether it is triggered by reconnection below or above the flux rope or by MHD instabilities \cite{Moore&Sterling2006}. To discriminate models with observations turns out extremely difficult due to the rapid development of eruptive structures, which is convoluted by the line-of-sight confusion in the optically thin corona and the projection effects of three-dimensional structures. 

Recent observation and modeling suggest that a similar mechanism involving a mini flux rope works for jets on much smaller scales \cite{Sterling2015,Wyper2017}. Moreover, theoretical progress in magnetic reconnection has demonstrated 
an inherently time-dependent, bursty picture of the current sheet to be characteristic of high-Reynolds-number plasmas \citep{Shibata&Tanuma2001,Loureiro&Udensky2016}. Mini flux ropes, also termed plasmoids, are continuously formed and ejected in a hierarchical, fractal-like fashion, which not only influences the reconnection rate but also enhances the particle acceleration efficiency in a Fermi-like process \cite{Drake2006,Nishizuka+Shibata2013}. Thus, flux ropes are key to understanding the diverse eruptive phenomena, but is there any physical connection between flux ropes involved in micro-scale dynamics and those in global-scale activities?

\section*{Results}
Here we present observations of how a CME ($\sim\,$10$^{11}$ cm; Figure~\ref{suppfig:cme}) builds up from a seed that forms via the coalescence of multiple plasmoids ($\sim\,$10$^8$ cm). The eruption occurs at the northeast solar limb at about 15:50 UT on 2013 May 13, producing an energetic X2.8-class flare and a fast, full-halo CME propagating at $\sim$1800\kms in the outer corona. The flare is observed in EUV by the Atmospheric Imaging Assembly (AIA \cite{lemen2012}; Methods) onboard the Solar Dynamics Observatory (SDO) and in hard X-rays (HXRs) by the Reuven Ramaty High-Energy Solar Spectroscopic Imager (RHESSI \cite{linrp2002}) and by the Gamma-ray Burst Monitor (GBM) onboard the Fermi Gamma-ray Space Telescope. AIA's six EUV passbands have distinctive temperature responses and cover a wide temperature range [0.5--30 MK], which allows us to reconstruct the temperature distribution of plasma emitting along the line of sight, known as differential emission measure (DEM; Methods). HXR characteristics during the impulsive phase of this flare has been studied in detail by \cite{Gou2017}. 
In the following, we analyze multi-wavelength observations to understand the initiation of the CME and its connection with magnetic reconnection and particle acceleration.

\begin{figure} 
	\centering
	\includegraphics[width=\textwidth]{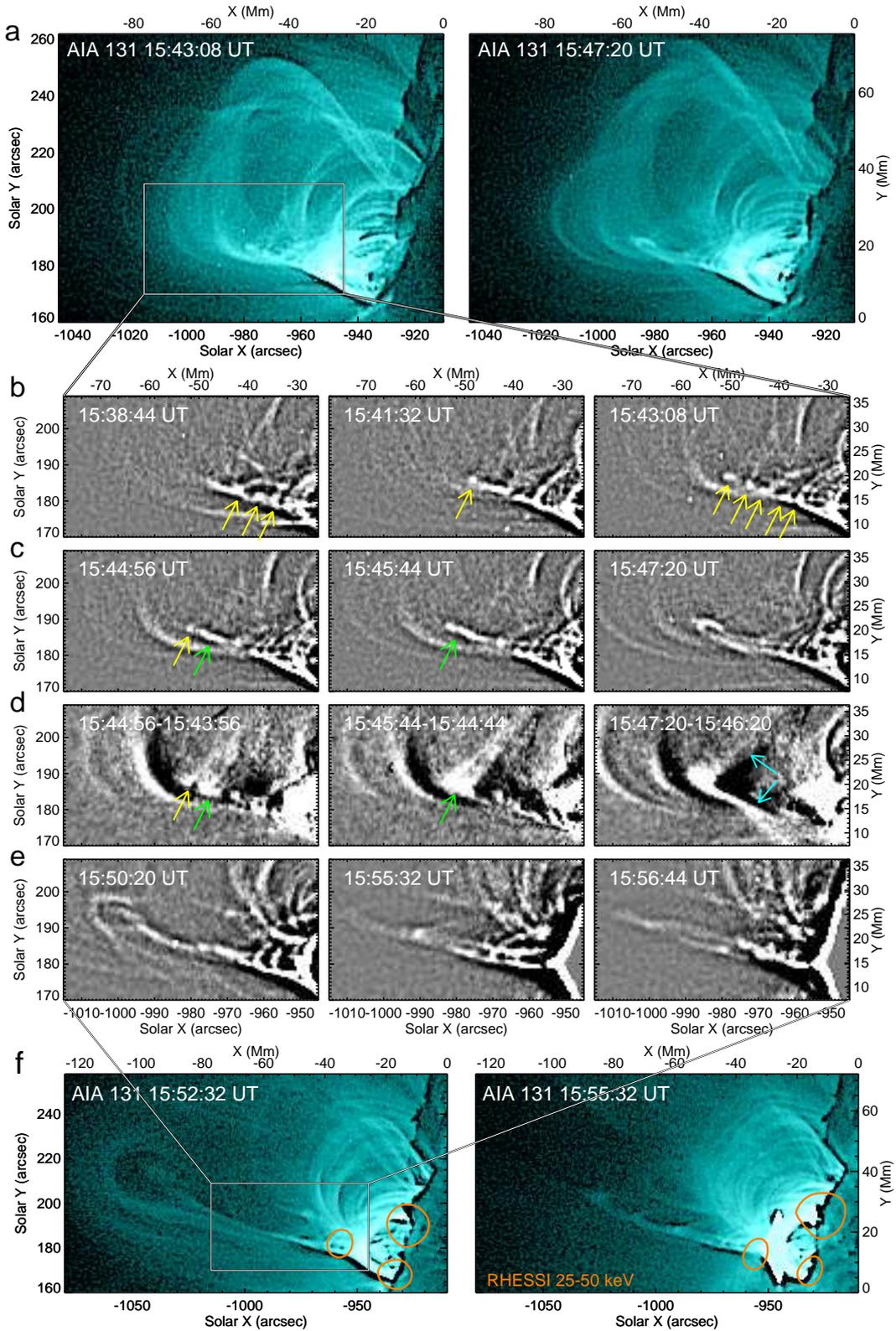}
	\caption{CME initiation and eruption. \textbf{a} and \textbf{f}: SDO/AIA 131~\AA\ enhanced images (Methods) showing the rising and expansion of the leading plasmoid at the upper tip of a linear extended feature, i.e., the current sheet. Orange contours (\textbf{f}) show RHESSI HXRs in the 25--50~keV band at 50\% of the maximum brightness. Plasmoids (marked by yellow arrows) in the current sheet are highlighted by unsharp masked (\textbf{b}, \textbf{c}, and \textbf{e}) and running-difference images (\textbf{d}). Coalescing plasmoids are marked by green arrows. The legs of the forming seed flux rope are marked by cyan arrows in difference images in \textbf{d}. 
	\label{fig:map}}
\end{figure}

\begin{figure} 
	\centering
	\includegraphics[width=\textwidth]{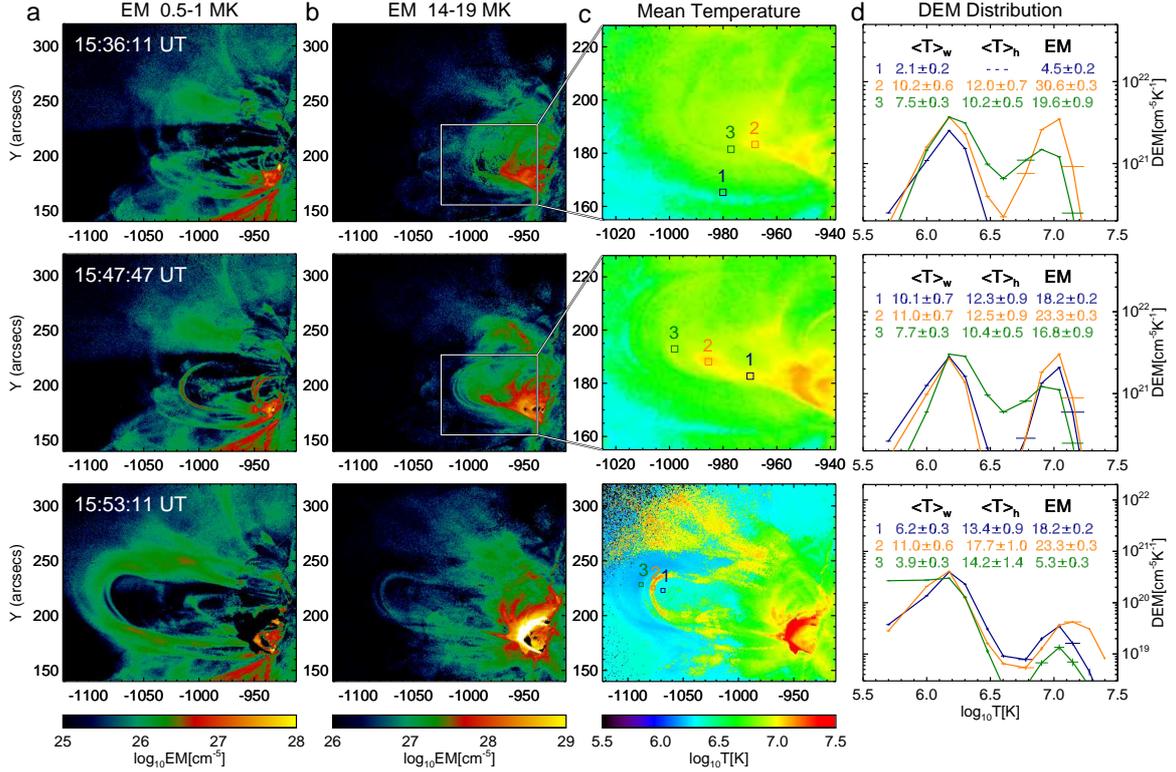}
	\caption{Plasma diagnostics with DEM analysis. \textbf{a.} EM maps at 0.5--1~MK showing cool coronal loops overlying the flux rope. \textbf{b.} EM maps at 14--19~MK showing prominent emission from the flux rope, current sheet and flare loops. The box in the top and middle indicates the field of view of the temperature maps in \textbf{c}. The EM maps in \textbf{a} and \textbf{b} are plotted in a logarithmic scale in units of cm$^{-5}$. Temperature maps in \textbf{c} are given over the whole temperature range (0.5--30~MK) and plotted in a logarithmic scale. \textbf{d.} DEM distribution in small regions of interest as shown in \textbf{c}. The size of sub-regions is 4$\times$4 pixels$^2$ in the top and middle panels and 6$\times$6 pixels$^2$ in the bottom panel. Calculated are the DEM-weighted temperatures $\langle T\rangle_w$ and $\langle T\rangle_h$ in units of MK (Methods) and EM in units of $10^{27}$ cm$^{-5}$ for each sub-regions. 
	\label{fig:dem}}
\end{figure}

The eruption results in a typical white-light CME with a diffuse outfront and a bright inner core (Figure~\ref{suppfig:cme}). The core bears similarity to the eruptive structure observed in the inner corona in EUV: a hollow ellipsoid connected to the top of flare loops by an extended linear feature of width $\sim\,$2$''$ during the impulsive phase of the flare (Figure~\ref{fig:map}f). The ellipsoid is only visible in AIA's hot passbands: best in 131~{\AA} (Fe XXI and XXIII), fairly in 94~{\AA} (Fe XVIII), and marginally in 335~{\AA} (Fe XVI; \suppmv~2). The linear feature is exclusively visible in 131~{\AA} (Figure~\ref{fig:map}) and emits at $>$10~MK as confirmed by the DEM analysis (Figure~\ref{fig:dem}; \suppmv~3). The ellipsoid also has a hot outer shell, though slightly cooler than the current sheet. Both the morphological and thermodynamic features fully agree with the standard model, in which hot plasma is expected to emit not only at the vertical current sheet and the flare loops, but in the outermost layer of the flux rope, where the magnetic field is newly reconnected \cite{Lin&Forbes2000,Lin2004}, while the inner layers of plasma have cooled. Thus, the ellipsoid is identified as the flux rope and the linear feature as the vertical current sheet, similar to previous studies \cite{Lin2005,Liu2010cs,Savage2010,Liu2013,Zhu2015}. 

\begin{figure} 
	\centering
	\includegraphics[width=\textwidth]{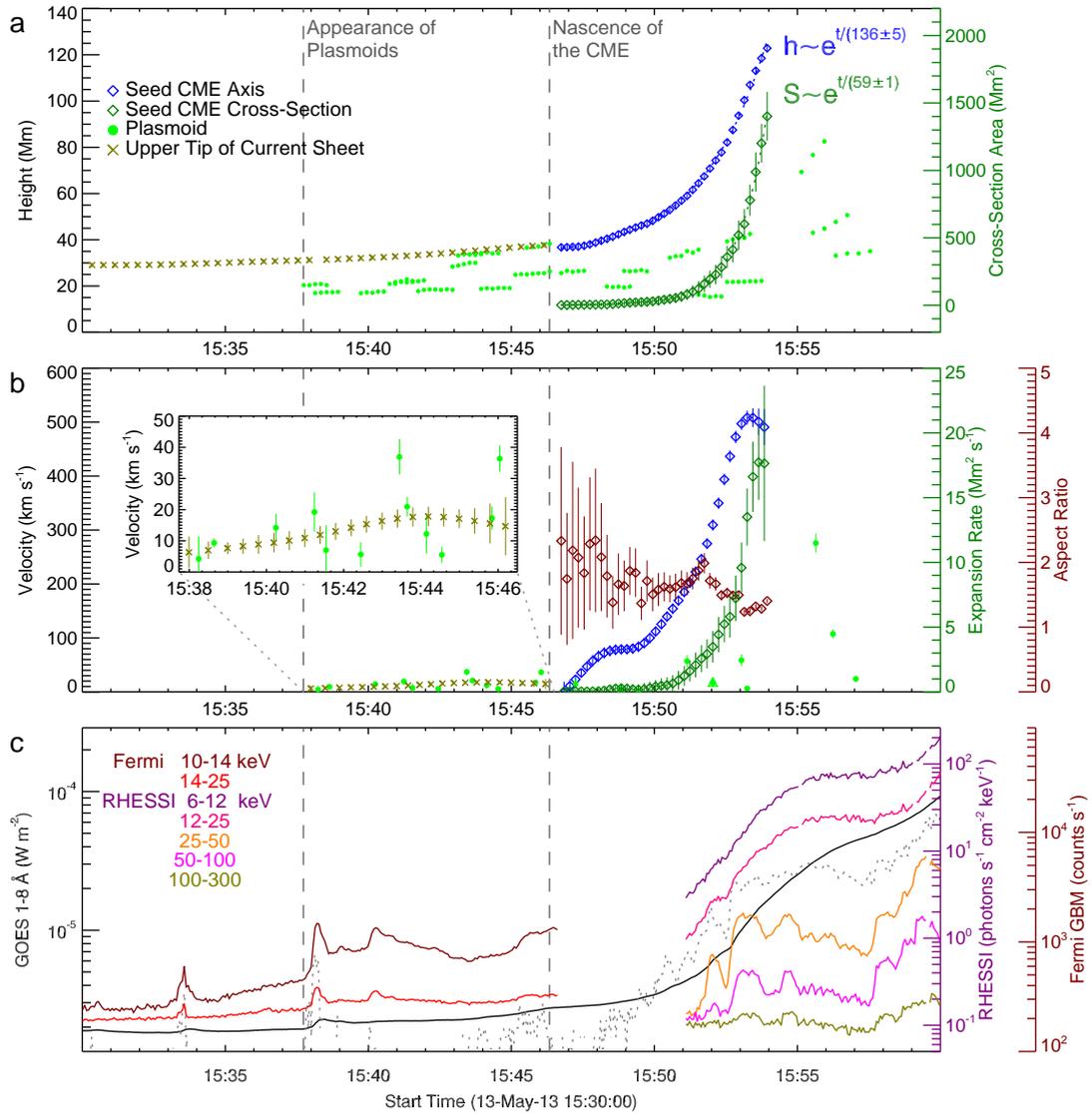}
	\caption{Kinematics of plasmoids in relation to X-ray emission. \textbf{a.} Projected heights of various structures (indicated by the legend) as scaled by the left y-axis, and the cross-section area of the flux rope as scaled by the right y-axis. Plasmoids are indicated by green dots. The flux rope evolves from the leading plasmoid at the upper tip of the current sheet (olive) and grows exponentially both in height (blue) and in cross-section area (dark green). \textbf{b.} Upward extension speed of the current sheet (olive), rising speed of the plasmoids (green) and of the flux-rope axis (blue), and expansion rate of the rope cross-section (dark green). The triangle at $\sim\,$15:52~UT shows the speed of a downward-moving plasmoid (see also Figure~\ref{suppfig:t-h}). The maroon diamonds show the aspect ratio of the ellipse fitting the flux rope. \textbf{c.} GOES 1--8~\AA\ SXR flux, HXR count rates (ending at $\sim\,$15:47~UT) recorded by the Gamma-ray Burst Monitor onboard the Fermi Gamma-ray Space Telescope, and HXR photon fluxes (starting from $\sim\,$15:51~UT) recorded by the Reuven Ramaty High-Energy Solar Spectroscopic Imager. The gray dotted line indicates the time derivative of GOES 1--8~\AA\ flux in an arbitrary unit, emulating HXRs because of the Neupert effect \cite{neupert1968}. The two vertical dashed lines indicate the time instances when plasmoids start to appear and when the seed flux rope starts to expand, respectively.  \label{fig:flux} }
\end{figure}

However, in the present case, the current sheet is not only present during the impulsive phase of the eruption, but evolves continuously from a shorter one visible already before the eruption. This pre-eruption current sheet is located immediately above an arcade of post-flare loops (Figure~\ref{fig:map}a), the remnant of a confined C5.3-class flare at 13:55 UT (\suppnote). With similar temperatures as high as 10 MK (Figure~\ref{fig:dem}[c,d]), both the post-flare arcade and the current sheet are located beneath a magnetically sheared arcade (\suppnote; \suppmv~4). This is inferred by performing stereoscopic triangulation on a low-lying flare loop and a high-lying coronal loop already stretched by the rising flux rope (Figure~\ref{suppfig:arcade}; Methods). As it extends upward slowly at $\sim\,$10~\kms (Figure~\ref{fig:flux}b), the current sheet is fragmented into multiple plasmoids of widths $\sim\,$2$''$ from about 15:38 UT onward (Figure~\ref{fig:map}b). With plasmoids showing up, Fermi GBM detects HXR bursts in 10--14 and 14--25~keV (Figure~\ref{fig:flux}c). At 15:41:32 UT, a leading plasmoid occupies the upper tip of the current sheet (Figure~\ref{fig:map}b). Underneath, a chain of at least four other plasmoids appear at 15:43:08 UT. Along the current sheet, the following plasmoids often move faster than (inset of Figure~\ref{fig:flux}b), and merge with, the leading plasmoid (Figure~\ref{fig:map}c; \suppmv~1). Two successive episodes of plasmoid coalescence can be seen at 15:44:56~UT and 15:45:44~UT (Figure~\ref{fig:map}[c,d]). This dynamic behavior is well established in numerical simulations (Methods; Figure~\ref{fig:plasmoids}).  As a result of plasmoid coalescence, from 15:47:20 UT onward, the current sheet is led by a larger plasmoid of width $\sim\,$4$''$, an ellipsoid characterized by hot plasma of 14--19~MK (Figure~\ref{fig:dem}b, \suppmv~3) . 

We interpret this ellipsoid as a `seed' flux rope because it continuously expands, eventually ballooning into the CME bubble, and while expanding it keeps a coherent shape, i.e., a hollow ellipsoid in AIA 131~{\AA} with an aspect ratio of about 1.5 (Figure~\ref{fig:flux}b). In the difference images that highlight the dynamic features, this seed flux rope exhibits two legs connecting to the surface, revealing its three-dimensional nature (Figure~\ref{fig:map}d; \suppmv~1). In contrast, the so-called plasma blobs reported previously are seemingly isolated features in current sheets observed in EUV images \cite{Liu2013,Zhu2015,takasao2016} or in white-light coronagraphs \cite{Lin2005,Liu2010}. This difference is probably due to instrumental resolution and sensitivity, because the present seed flux rope is apparently the largest plasmoid ever reported in EUV. 

While expanding, the flux rope becomes a hollow ellipsoid, i.e., depressed in 131~{\AA} emission in its center, but as a whole it becomes a dark `cavity' in cool passbands such as 171~{\AA} (Figure~\ref{fig:cartoon} and \suppmv~2). The absence of plasma emission inside the rope is well expected for a twofold reason: first, these plasmas are brought into the rope by earlier reconnections at the current sheet, and have since cooled down via conduction, radiation, and expansion, while the hot `rim' is produced by the most recent reconnections; second, the dominant magnetic pressure inside the rope tends to squeeze out plasma, so that the total pressure, magnetic plus plasma pressure, is balanced with the surroundings. 

As soon as it forms at about 15:47 UT, the seed flux rope starts to rise at a speed of tens of kilometers per second (Figure~\ref{fig:flux}). Initially increasing with time at $\sim\,$1~km~s$^{-2}$, the rising speed temporarily plateaus at about 15:49 UT at $\sim\,$80~\kms, short of the sound speed in the corona. From about 15:50 UT, the speed quickly increases at $\sim\,$2.5~km~s$^{-2}$ and reaches at about 15:53 UT a peak velocity of $\sim$530\kms, comparable to the Alfv\'{e}n speed in the inner corona. Plasmoids still appear intermittently in the current sheet at this stage (Figures~\ref{fig:map} and \ref{fig:flux}, \suppmv~1), and they generally have larger sizes than in the early stage and move at faster speeds ranging from tens \kms up to 300 \kms. In tandem with the enhanced acceleration, the rope's cross-section area expands exponentially from 15:50~UT onward (Figure~\ref{fig:flux}a), which is associated with a rapid increase in SXR and HXR fluxes, suggesting an increase in reconnection rate, which might be induced by plasmoid ejection \cite{Schumacher&Kliem1996,Shibata&Tanuma2001}. The HXRs in the nonthermal range ($>25$ keV) emit with intermittent, spiky profiles (Figure~\ref{fig:flux}c), which is generally taken as a signature of the reconnection-related electric field that rapidly varies with time and/or space \cite{Kliem&al2000}, therefore modulating the particle acceleration. 

Meanwhile, the expanding and rising flux rope starts to stretch and compress the overlying loops (Figure~\ref{suppfig:t-h}), 
which are visible both in 131~{\AA} due to the \ion{Fe}{8} line blend and in cooler passbands, namely, AIA 171, 193 and 211~{\AA} (Figure~\ref{fig:dem}, \suppmv~2). The legs of these loops are first stretched longer and longer, and then, as the flux rope expands, their top part becomes wider than the lower part, exhibiting an $\Omega$ shape (Figure~\ref{fig:cartoon}). At $\sim\,$15:53~UT when the rope's rising speed peaks over 500~\kms, such an $\Omega$-shaped, thin layer appears in the EM map of hot plasma (14--19~MK) as well as the map of mean temperature (Figure~\ref{fig:dem}, \suppmv~3), apparently separating the overlying loops from the flux rope. Detailed DEM analysis (Methods; Figure~\ref{fig:dem}d) shows that it is significantly hotter and denser than both the overlying loops and the flux rope, and hence interpreted as a quasi-separatrix layer (QSL) that wraps around the flux rope to separate the twisted from untwisted field \cite{liu2016}. A flux rope's QSL boundary is known as a preferential site for current concentration \cite{Demoulin2006}. On the solar surface, this would correspond to the boundary of the rope's feet, which has recently been observed as a pair of irregular bright rings expanding from points during the flare impulsive phase \cite{Wang2017}, indicating a flux-rope formation process similar as reported here. Plasma compression is conducive to current steepening and dissipation in this QSL, which explains the elevated temperature and density.

\section*{Discussion}
\begin{figure} 
	\centering
	\includegraphics[width=\textwidth]{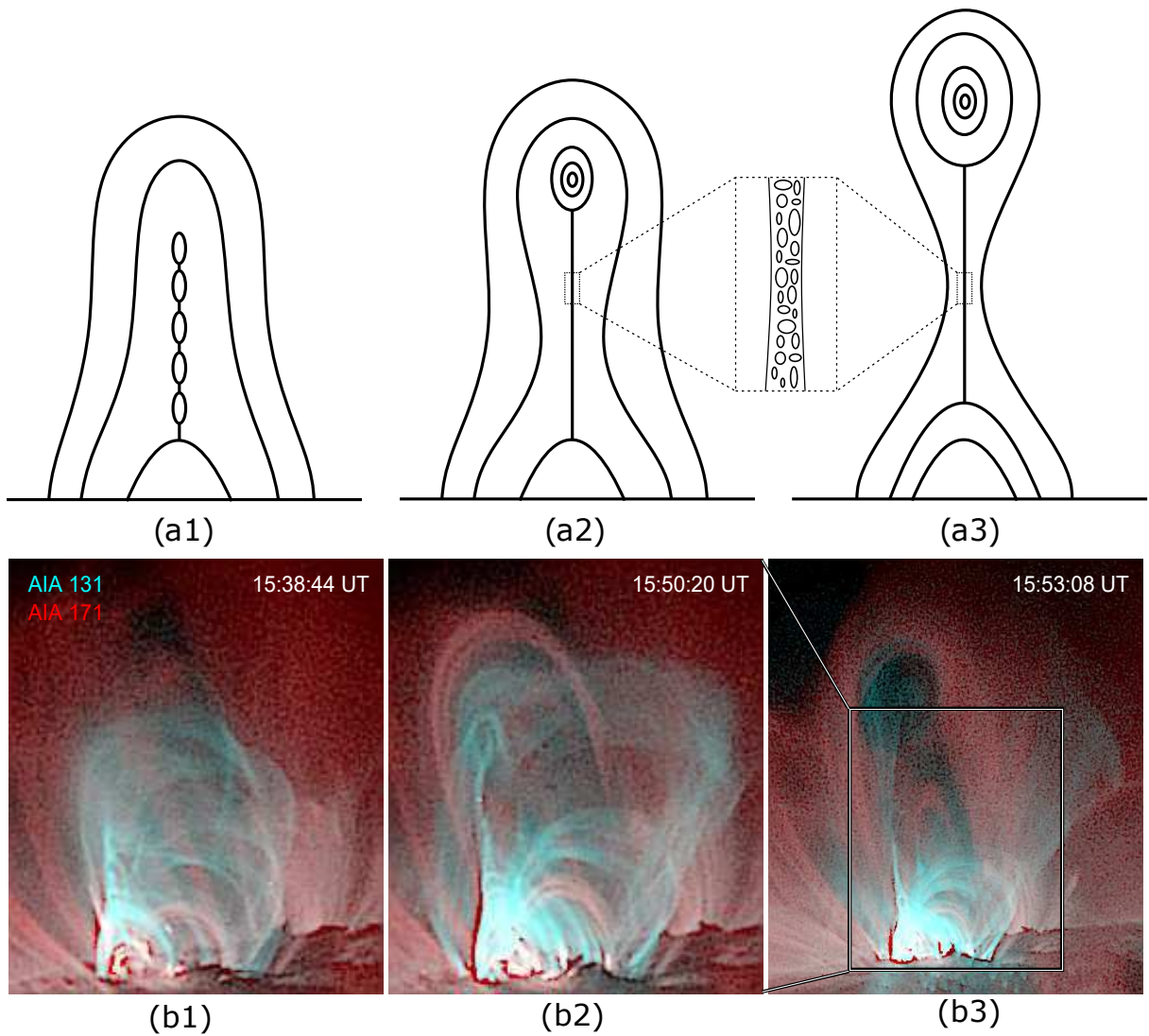}
	\caption{Schematic view of the CME initiation process, matched by observations. \textbf{a1--a3} A vertical current sheet underneath a magnetically sheared arcade breaks up into multiple plasmoids. The ejection and coalescence of plasmoids initiate a seed flux rope, which rises and stretches the overlying field. Consequently, fast reconnection is induced at the current sheet, which possesses plasmoids of various scales, as illustrated by the inset. \textbf{b1--b3} Composite images of AIA 131~\AA\ ($\sim\,$10~MK; cyan) and 171~\AA\ ($\sim\,$0.7~MK; red). The AIA images have been rotated 90 degree clockwise. Note the field of view in \textbf{b1} and \textbf{b2} is smaller than that in \textbf{b3}. \label{fig:cartoon}}
\end{figure}

Illustrated and matched by observations in Figure~\ref{fig:cartoon} is the process of the CME eruption in a two-dimensional cross section: Initially a vertical current sheet exists underneath a magnetically sheared arcade (see \suppnote for its formation). The current sheet extends as magnetic energy builds up slowly in the corona and breaks up into plasmoids when its length exceeds the critical wavelength for the tearing-mode instability \cite{Furth&al1963} (Figure~\ref{fig:cartoon}(a1)). In three dimensions, this corresponds to the transformation of sheared into twisted field lines via magnetic reconnection  \cite{Moore2001,Liu2010tc}. The plasmoids are then propelled to move along the current sheet by the magnetic tension force, while neighboring plasmoids merge into larger ones due to the coalescence instability \cite{Pritchett&Wu1979}.

Upward moving plasmoids eventually merge with the leading plasmoid at the upper tip of the current sheet. A coherent flux rope hence starts to form (Figure~\ref{fig:cartoon}(a2)). Because of its hoop force \cite{Kliem&Torok2006} and the upward reconnection outflows, the rope keeps rising, stretching the overlying field and driving faster plasma inflow into the current sheet, owing to the conservation of mass, therefore thinning the current sheet and enhancing the reconnection rate. Overlying field is now reconnecting at the current sheet, adding magnetic flux to the flux rope (Figure~\ref{fig:cartoon}(a3)); more flux makes the rope rise faster, which in turn leads to faster reconnection rate. At this point, a positive feedback is established \cite{Lin2004,Temmer2010}, and 

the flux rope can grow into a runaway CME bubble. The close coupling between the flux-rope eruption and particle acceleration strongly suggests that while the plasmoids are building up into the CME, they are simultaneously cascading into smaller and smaller scales (illustrated by the inset of Figure~\ref{fig:cartoon}(a2 and a3)) in a fractal fashion down to ion and electron kinetic scales at which the energetic particles are actually accelerated \cite{Shibata&Tanuma2001,Loureiro&Udensky2016}. The observed plasmoids at a meso-scale of 10$^8$ cm thus bridge the macro- (10$^{11}$ cm) and micro-scale (10$^4$ cm) flux ropes across a hierarchical spectrum. 

If the overlying constraining field is strong enough, then the eruption can be  confined, which is also termed failed eruption \cite{Torok&Kliem2005,Kliem&Torok2006,Amari2018}. In that case, the flux rope may temporarily settle down, and the bottom of its helical field lines can support a prominence made of relatively cool and dense plasma \cite{Mackay2010}. However, as the current sheet continues to spawn plasmoids and the plasmoids continue to merge into the flux rope, the accumulated flux in the rope may eventually reach the tipping point of eruption \cite{Zhang2014}. Further, whenever open field is accessible to the plasmoids, a jet ensues instead of a CME \cite{Shibata1999,Sterling2015,Wyper2017}. 

\section*{Methods}
\subsection*{SDO Data and DEM Analysis} 

SDO/AIA provides full-disk observations of the Sun at high spatial ($1''.2$) and temporal (12s) resolution around the clock. AIA's six EUV passbands, i.e., 131~{\AA} (\ion{Fe}{21} for flare, peak response temperature $\log T = 7.05$; \ion{Fe}{8} for AR, $\log T = 5.6$ \cite{ODwyer2010}), 94~{\AA} (\ion{Fe}{18}, $\log T = 6.85$), 335~{\AA} (\ion{Fe}{16}, $\log T = 6.45$), 211~{\AA} (\ion{Fe}{14}, $\log T = 6.3$), 193~{\AA} (\ion{Fe}{24} for flare, $\log T = 7.25$; \ion{Fe}{12} for AR, $\log T = 6.2$), and 171~{\AA} (\ion{Fe}{9}, $\log T = 5.85$), are used to calculate the differential emission measure (DEM), which characterizes the amount of optically thin plasma at a specific temperature along the line of sight. We adopted the regularized inversion code developed by Hannah\&Kontar \cite{Hannah&Kontar2012} and considered the DEM solutions of relative uncertainties $\leqslant$30\% with temperature bins $\log T\leqslant 0.5$. The emission measure (EM) is obtained by integrating DEMs over the temperature ranges $\log T = 5.5 - 7.5$ and the DEM-weighted mean temperature is calculated by $\langle T\rangle=\frac{\sum DEM(T) \times T\Delta T}{\sum DEM(T) \Delta T}$. To characterize flaring plasma, we calculated $\langle T\rangle_h$ for temperatures above 4~MK ($\log T \approx 6.6$) besides $\langle T\rangle_w$ which employs the whole temperature range, considering that plasma below 4~MK is mainly contributed by the background corona \cite{Gou2015}. 

\subsection*{Unsharp Masking}

We applied unsharp masking to SDO/AIA 131~\AA\ images to highlight fine structures like plasmoids. First, one generates a pseudo background by smoothing the original image with a box-car, and then obtains the unsharp masked image, the residual of subtracting the background from the original, or the enhanced image by adding the original by a factor back to the residual. We adopted a smoothing window of $5\times5$~pixels ($3''\times3''$) for the early stage evolution (15:30--15:50~UT) and $7\times7$~pixels later on for optimal effects. 

\subsection*{Kinematics of Plasmoids and Flux Rope}

We visually identified plasmoids that persist over a few frames in a series of AIA 131~\AA\ unsharp masked images, obtained their projected heights above the solar limb, assuming an measurement error of 2~pixels ($1''.2$), and estimated the speed by linearly fitting the time-height profile of each plasmoid (Figure~\ref{fig:flux}). Similarly we obtained the time-height profile of the current sheet tip and because of its smooth, continuous extending we derived the speed by numerical derivatives using the IDL procedure \texttt{DERIV.pro}. 

We fitted the expanding hollow ellipsoid in AIA 131~\AA\ by an ellipse, assigning a conservative 4 pixels for the uncertainties of the fitting parameters, i.e., the center and the two semi-axes. We considered the center as the axis of the flux rope, and derived the rising speed of the axis and the expanding rate of the ellipse area by numerical derivatives.

We have also constructed time-distance diagrams by taking slices off the original or running-difference images oriented along the current sheet (dotted line in the inset of Figure~\ref{suppfig:t-h}a) and then stacking them up chronologically. Structures moving along the slit leave clear tracks on these diagrams. 

\subsection*{3D Perspective of the Eruption}
The active region of interest, NOAA AR 11748, is located on the disk as seen from the ``Behind'' satellite of the Solar Terrestrial Relations Observatory (STEREO-B) \cite{Kaiser2008}, which is about 141.6$^\circ$ behind Earth on its ecliptic orbit on 13 May 2013. The 195~{\AA} channel of the Extreme Ultraviolet Imager (EUVI) onboard STEREO has a similar temperature response as SDO/AIA 193~\AA, which allows us to perform stereoscopic triangulation on a low-lying flare loop and a high-lying coronal loop already stretched by the rising flux rope at $\sim\,$15:46 UT (Figure~\ref{suppfig:arcade}(a and b)). With their true heights being recovered, the reconstructed loops are projected above a photospheric B$_z$ map observed taken 6 days later when the active region is located near the disk center (Figure~\ref{suppfig:arcade}(c)). One can see that both the flare loop and the overlying loop are highly sheared with respect to the polarity inversion line, which is mostly east-west oriented. The space in between is supposedly occupied by the flux rope (c.f. Figure~\ref{fig:cartoon}(b2)). Both loops are anchored in the vicinity of conjugate coronal dimmings in EUVI 195~{\AA} (Figure~\ref{suppfig:arcade}(d)), where coronal mass escapes along the CME field into the interplanetary space, therefore mapping the CME footpoints. 

\subsection*{Plasmoid dynamics in CME current sheet} \label{suppnote:plasmoids}
We have studied the formation and dynamics of multiple plasmoids in a CME simulation which provides a realistic 3D setting for these processes, previously realized only in ref\cite{Nishida&al2013}, where the formation of multiple plasmoids was also found but their coalescence was not studied. Our simulation is very similar to a less resolved simulation in \cite{Schrijver&al2008} (see their Figure~7), here with a slightly higher growth rate of the initial torus-unstable flux rope equilibrium \cite{Titov&Demoulin1999}. Upon erupting, the flux rope spawns a vertical current sheet as in the standard model \cite{Lin&Forbes2000} and sets up the inflows into the current sheet. These initiate the reconnection which is allowed by the numerical diffusion of the field in the ideal MHD simulation. Careful comparison of such simulations with well-observed solar eruptions has shown that they reproduce the observed overall reconnection rate quite accurately, within a factor of two \citep{Kliem&al2013, Hassanin&Kliem2016}, because the reconnection in these events is driven by the large-scale flux rope instability, whose flows regulate the reconnection rate. When the current sheet has lengthened to an aspect ratio of $\sim10^2$, multiple X- and O-lines begin to form. These always immediately tend to merge into larger O-type structures, like plasmoids, which here are small seed flux ropes extending up to $\sim35$ current sheet half widths in the horizontal (current) direction. At most times, there are several such 3D plasmoids of different size in the current sheet. All of them are eventually ejected with the large-scale reconnection outflow, either upward into the erupted flux rope, which is topologically equivalent to the merging of the observed plasmoids with the leading one, or downward into the growing flare loop arcade. Dynamic plasmoids are seen as long as the run is continued, see Figure~\ref{suppfig:plasmoids} and Supplementary Movie~5. The simulation reveals dynamic plasmoids, basically similar to the behavior seen in many 2D simulations, when the current sheet is long enough \cite{Loureiro&Udensky2016}. Therefore, we expect the same behavior in a vertical current sheet formed prior to the onset of eruption and more slowly driven by a photospheric process as expected for the present event, once the current sheet has reached a sufficient height, i.e., aspect ratio. This, however, remains to be verified by a future simulation study.


\clearpage
\begin{addendum}
	\item [Acknowledgments]  T.G. and R.L. are supported by NSFC 41474151, 41774150, and 41761134088. R.L. also acknowledges the	Thousand Young Talents Program of China. Y.W. acknowledges support by NSFC 41774178 and 41574165. B.K. acknowledges support by the DFG and NSFC through the collaborative grant KL 817/8-1/NSFC. A.V. acknowledges support by the Austrian Science Fund (FWF) P27292-N20. This work is also supported by NSFC 41421063, CAS Key Research Program of Frontier Sciences QYZDB-SSW-DQC015, and the fundamental research funds for the central universities.
	\item [Competing Interests] The authors declare that they have no competing financial interests.
	\item [Contributions] R.L. interpreted the data and wrote the manuscript. T.G. processed and analyzed the AIA data. B.K. performed the MHD simulation. B.K., Y.W., and A.V. participated in the discussions and made contributions to finalize the manuscript.		
	\item[Data Availability] All data needed to evaluate the conclusions in the paper are present in the paper and/or the Supplementary Materials. Additional data available from authors upon request.
\end{addendum}

\clearpage
\spacing{1.0}
\small

\clearpage
\spacing{1}
\renewcommand{\thefigure}{S\arabic{figure}} 
\setcounter{figure}{0}
\section*{Supplementary Figures}
\begin{figure}[htb]
	\centering
	\includegraphics[width=\textwidth]{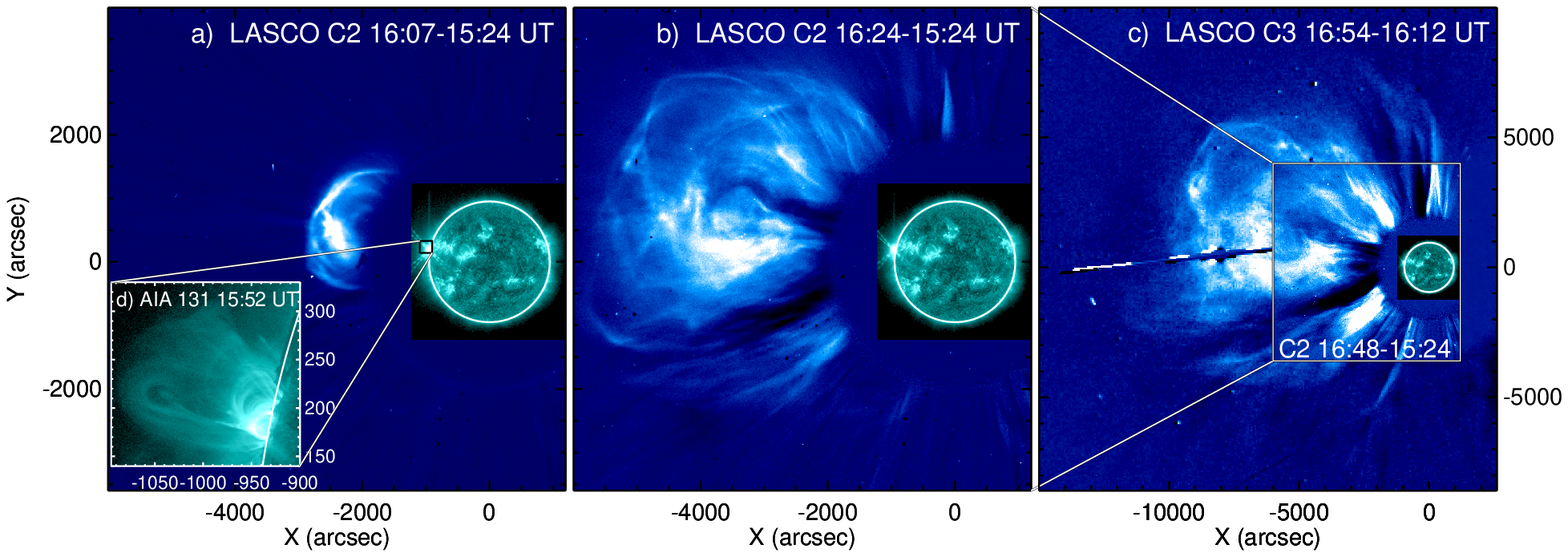}
	\caption{Full-halo CME observed by the Large Angle and Spectrometric Coronagraph onboard Solar and Heliospheric Observatory. LASCO's C2 coronagraph has a smaller field of view (2.2--7 solar radii) than C3 (6--30 solar radii). A background image is subtracted from each coronagraph images. The CME's core structure (\textbf{b}) displays similar morphology as the expanding ellipsoid observed in EUV in the inner corona (\textbf{d}). \label{suppfig:cme}}
\end{figure}

\begin{figure}[htb]
	\centering
	\includegraphics[width=\textwidth]{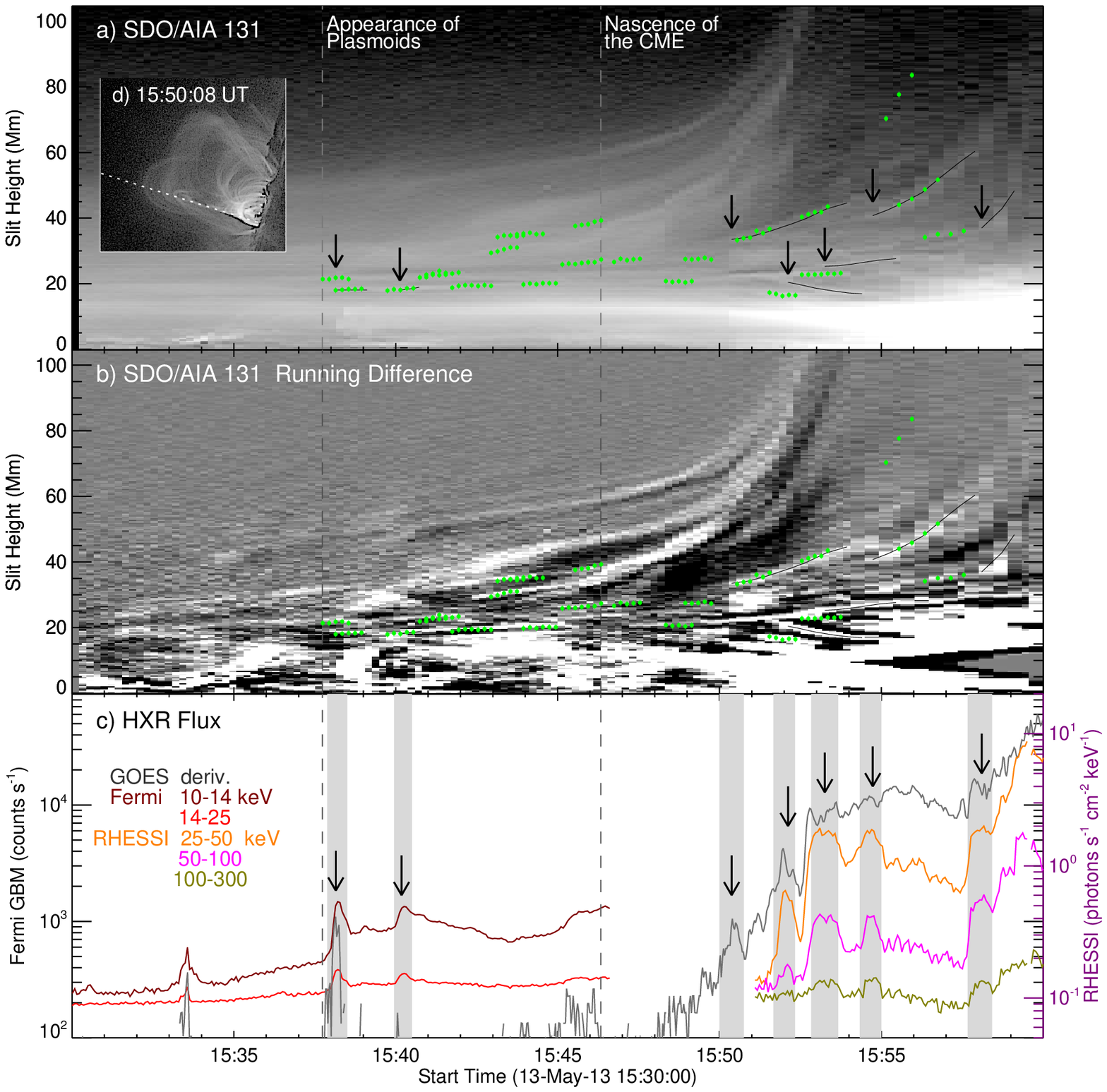}
	\caption{Dynamic evolution of the plasmoids. A slit of width 8~pixels ($\sim4''.8$) is placed above the solar limb along the current sheet, indicated by the dotted line in the inset of Panel \textbf{a}, to construct the time-distance diagrams using 131~\AA\ images enhanced by unsharp masking (\textbf{a}) and 131~\AA\ running-difference images (\textbf{b}). The identified plasmoids in Figure~\ref{fig:flux} are re-plotted in the time-distance diagrams as green dots. A few tracks left by moving plasmoids are delineated by black curves. During $\sim\,$15:50--15:55~UT, some plasmoids at the height of $\sim\,$20~Mm move downward to merge into the flare loops (\suppmv~1). \textbf{c.} HXR count rates recorded by the Gamma-ray Burst Monitor onboard the Fermi Gamma-ray Space Telescope, and HXR photon fluxes recorded by the Reuven Ramaty High-Energy Solar Spectroscopic Imager. The gray line plots the time derivative of GOES 1--8~\AA\ flux in an arbitrary unit. Arrows mark plasmoids whose appearance are temporally associated with HXR bursts (gray bars). \label{suppfig:t-h}} 
\end{figure}

\begin{figure}[htb]
	\centering
	\includegraphics[width=\textwidth]{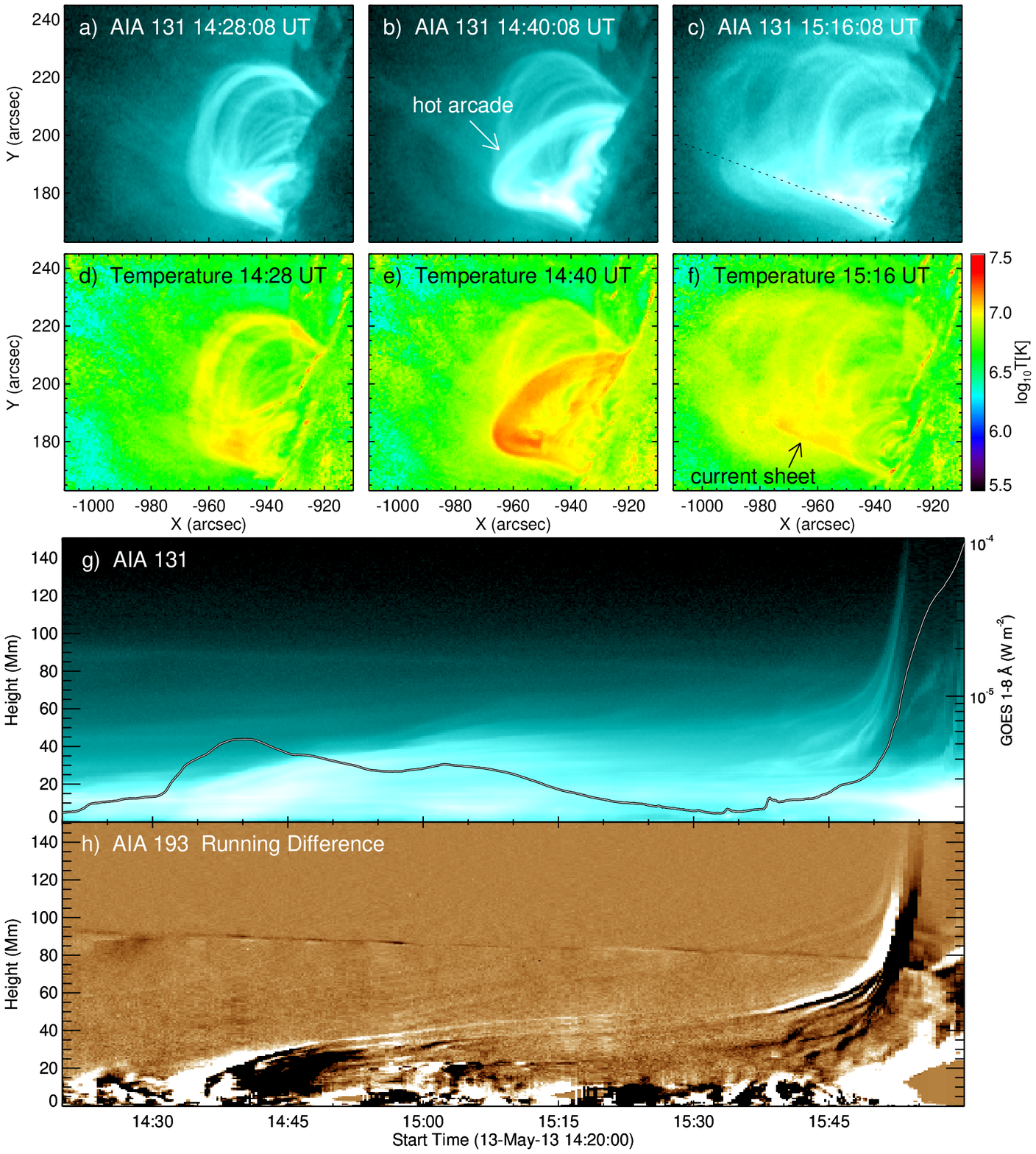}
	\caption{Pre-eruptive evolution of the active region. Panels \textbf{a--c} show snapshots of SDO/AIA 131~\AA\ observations, with corresponding maps of DEM-weighted mean temperature in \textbf{d--f}. In \textbf{b} and \textbf{e} a hot arcade emerges in the active region. After the arcade cools down (\textbf{c} and \textbf{f}), a hot linear feature, which is vertically oriented above flare loops, is interpreted as a current sheet. \textbf{g} and \textbf{h} plot time-distance diagrams seen through a slit along the current sheet (dotted line in \textbf{c}) in 131 and 193~{\AA}, respectively. Overplotted in \textbf{g} is GOES 1--8~\AA\ flux as scaled by the right $y$-axis. \label{suppfig:arcade}}
\end{figure}

\begin{figure}[htb]
	\centering
	\includegraphics[width=\textwidth]{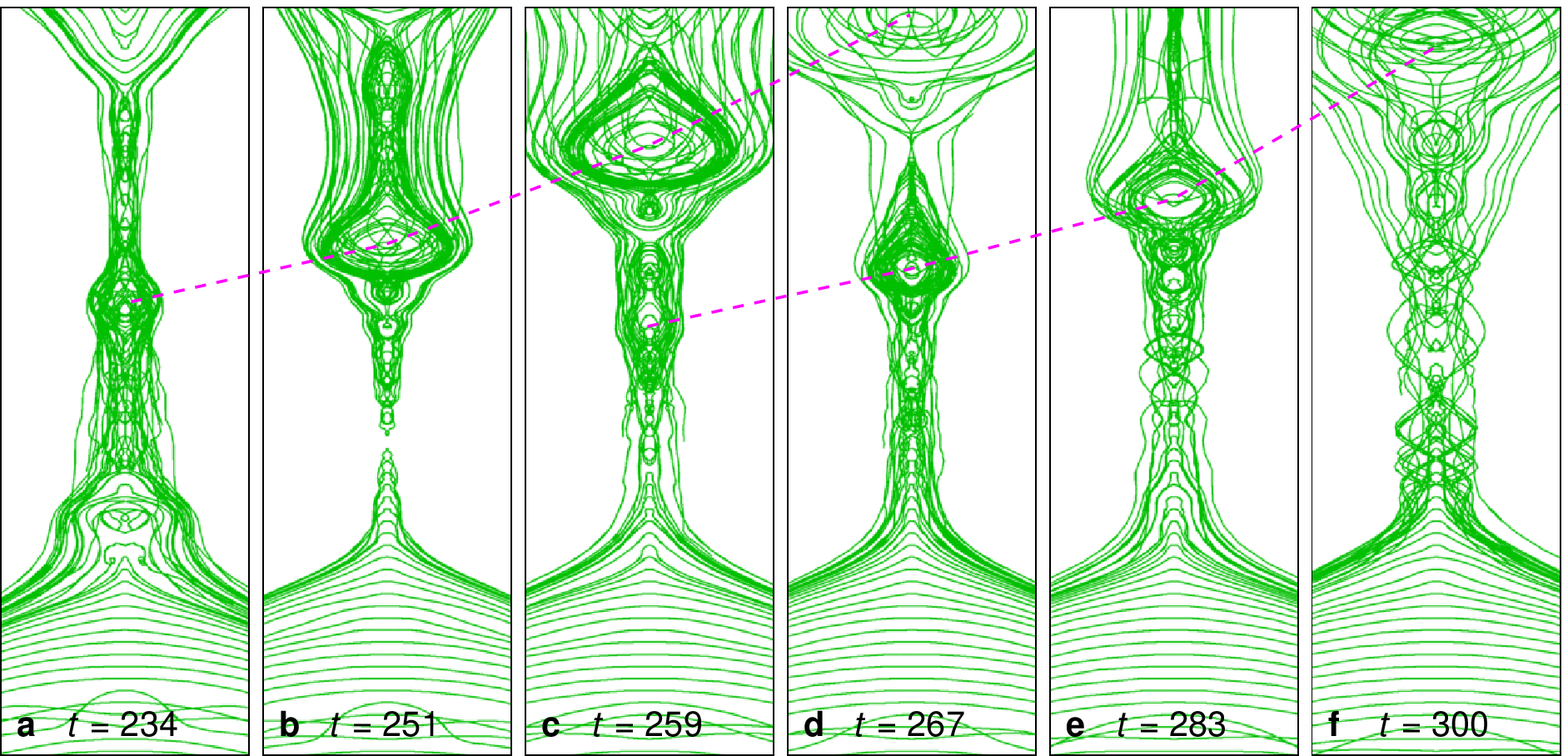}
	\caption{3D MHD simulation of multiple plasmoid formation and ejection in the vertical current sheet spawned by an erupting flux rope. Magnetic field lines, traced from equidistant points on the $z$ axis, show the dynamic current sheet between the lower edge of the erupting and strongly expanding flux rope and the slowly growing arcade of reconnected field lines, which represent the arcade of flare loops. The displayed volume, $|x|<2/7$, $|y|<0.25$, $2<z<8$, is stretched by a factor 7 in the horizontal direction for clarity of the plasmoids, two of which are traced by dashed lines. The short extent in $y$ direction is chosen because many of the small new flux ropes, seen here as plasmoids, bend upward or downward out of the plane. Time is given in units of the Alfv\'en time based on the initial flux rope height ($z=1$) and the peak Alfv\'en velocity in the initial configuration. \label{suppfig:plasmoids}}
\end{figure}


\clearpage
\section*{Supplementary Movies}
\begin{addendum}
	\item[Supplementary Movie 1] Plasmoid coalescence developing into a seed CME. The bottom left panel shows AIA 131~\AA\ images enhanced by unsharp masking, and the inset highlights the current sheet region and plasmoids. The bottom right panel shows the corresponding running-difference images.
	
	\item[Supplementary Movie 2] The eruption observed by SDO/AIA's six EUV channels. Different features show up in different passbands: the plasmoids and current sheet in 131~\AA, the flux rope in 131, 94 and 335~\AA, the overlying coronal loops in 211, 193 and 171~\AA.
	
	\item[Supplementary Movie 3] Temperature structure of the eruption revealed by the DEM analysis.
	
	\item[Supplementary Movie 4] Formation and evolution of the sheared arcade during the earlier confined C5.3-class flare.
	
	\item[Supplementary Movie 5] Plasmoid dynamics in the vertical current sheet of a CME simulation.
\end{addendum}

\spacing{1.5}
\section*{Supplementary Notes}
While small-scale current sheets exist ubiquitously in the corona because of the continuous shuffling and intermixing of field-line footpoints in the photospheric convection \cite{Parker1988}, a large-scale current sheet can be produced in the corona in numerous ways besides the well-known heliospheric current sheet, e.g., when an emerging flux collides with the coronal field \cite{Heyvaerts1977}, when a moving bipole bumps into surrounding magnetic structures \cite{Priest1994}, when a shearing magnetic arcade collapses under a hypothesized global resistive MHD instability \cite{Mikic+Linker1994} or interacts with an overlying magnetic null point \cite{Karpen2012}, or, when a quadrupole field is arranged in certain ways \cite{Liu2016SR}. 

The formation of the pre-existing current sheet in the present study can be traced back to a C5.3-class confined flare that starts at 13:55 UT and peaks at 14:40~UT, one hour before the X2.8-class flare. An arcade of loops rises and expands into the corona at $\sim\,$8~\kms from $\sim\,$14:30~UT, in tandem with the appearance of flare loops underneath (Figure~\ref{suppfig:arcade}; \suppmv~4). This arcade resembles the so-called ``hot channel''\cite{Zhang2012}, as being heated to temperatures over $\sim\,$10~MK, but unlike those interpreted as the proxy of flux ropes \cite[e.g.,][]{Zhang2012,Patsourakos2013}, it shows no signature of twist. The structure fades away as it cools down, from SDO/AIA 131 and 193~{\AA}, as well as from an alternative perspective in STEREO-B/EUVI 195~{\AA} (Figuree~\ref{suppfig:2view}). We suggest that this hot arcade illuminates flux that erupts to produce this confined C5.3 flare, whose associated reconnection heats the plasma involved. As the expanding arcade becomes faint a bright linear feature appears only in 131~\AA\ from $\sim$15:00~UT, which is as hot as the flare loops underneath ($\geqslant$10~MK), therefore being interpreted as the current sheet. The sustained heating is attributed to slow magnetic reconnection ongoing at the current sheet, until fast reconnection is triggered at the onset of the imminent X2.8 flare.

\end{document}